\documentclass{svmult}

\usepackage{makeidx}         
\usepackage{graphicx}        
\usepackage{multicol}        
\usepackage[bottom]{footmisc}

\makeindex             

\begin{document}
\title*{A Finite Baryon Density Algorithm}
\author{Keh-Fei Liu}
\institute{Dept. of Physics and Astronomy, University of Kentucky, Lexington,
KY 40506 USA
\texttt{liu@pa.uky.edu}~\footnote{Talk given at
`Third International Workshop on QCD and Numerical Analysis', Edinburgh, June 2003,
 preprint UK/03-18}}

%
%
\maketitle

   I will review the progress toward a finite baryon density algorithm in
the canonical ensemble approach which entails particle number projection from
the fermion determinant. These include an efficient Pad\'{e}-Z$_2$
stochastic estimator of the $Tr \log$ of the fermion matrix and a Noisy Monte
Carlo update to accommodate unbiased estimate of the probability. Finally, I
will propose a Hybrid Noisy Monte Carlo algorithm to reduce the large
fluctuation in the estimated $Tr \log$ due to the gauge field which should 
improve the acceptance rate. Other application such as treating $u$ and 
$d$ as two separate flavors is discussed.

\section{Introduction}
\label{introduction}

   Finite density is a subject of keen interest in a variety of topics, such as
nuclear equation of state, neutron stars, quark gluon plasma and color superconductivity 
in nuclear physics and astrophysics, and high temperature superconductors in condensed matter 
physics. Despite recent advance with small chemical potential at finite temperature~\cite{kat03},
the grand canonical approach with chemical potential remains a problem
for finite density at zero temperature. 

   The difficulty with the finite chemical potential in lattice QCD stems from the 
infamous sign problem which impedes important sampling with positive probability. 
The partition function for the grand canonical ensemble 
is represented by the Euclidean path-integral
\begin{equation}   \label{ZGC}
Z_{GC}(\mu) = \int {\cal{D}} U det M[U,\mu] e^{-S_g[U]},
\end{equation}
where the fermion fields with fermion matrix $M$ has been integrated to give the
determinant. $U$ is the gauge link variable and $S_g$ is the gauge action.
The chemical potential is introduced to the quark action with the $e^{\mu a}$ factor
in the time-forward hopping term and $e^{ - \mu a}$ in the time-backward hopping
term. Here $a$ is the lattice spacing. However, this causes the fermion action
to be non-$\gamma_5$-Hermitian, i.e. $\gamma_5 M \gamma_5 \neq M^{\dagger}$. As a result, 
the fermion determinant $det M[U]$ is complex that leads to the sign problem.

\section{Finite Chemical Potential}
\label{finite_chemical}

   There are several approaches to avoid the sign problem. 
 It was proposed by the Glasgow group~\cite{bmk98} that
the sign problem can be circumvented based on the expansion
of the grand canonical partition function in powers of the
fugacity variable $e^{\mu/T}$,
\begin{equation}   \label{zgc}
Z_{GC} (\mu/T, T, V) = \sum_{B = - 3V}^{B = 3V} e^{\mu/T\, B} Z_B(T,V),
\end{equation}
where $Z_B$ is the canonical partition function for the baryon sector with
baryon number $B$. $Z_{GC}$ is calculated with reweighting of the
fermion determinant
Since $Z_{GC}(\mu/T, T, V)$ is calculated with reweighting based on the gauge 
configuration with $\mu = 0$, it avoids the sign problem. However, this does not work, 
except perhaps for small $\mu$ near the finite temperature phase transition.
We will dwell on this later in Sec. {\ref{finite_baryon}. This is caused by
the `overlap problem'~\cite{alf99} where the important samples of configurations
in the $\mu = 0$ simulation has exponentially small overlap with those
relevant for the finite density. To alleviate the overlap problem, a reweighting
in multi-parameter space is proposed~\cite{fk02} and has been applied to study the end 
point in the T-$\mu$ phase diagram. In this case, the Monte Carlo simulation is carried
out where the parameters in the set $\alpha_0$ include $\mu = 0$ and $\beta_c$ which
corresponds to the phase transition at temperature $T_c$. The parameter
set $\alpha$ in the reweighted measure include  $mu \neq 0$ and
an adjusted $\beta$ in the gauge action. The new $\beta$ is determined 
from the Lee-Yang zeros so that one is following the transition line
in the T-$\mu$ plane and the large change in the determinant ratio
in the reweighting is compensated by the change in the gauge action to
ensure reasonable overlap. This is shown to work to locate the transition line
from $\mu =0$ and $ T = T_c$ down to the critical point on the
$4^4$ and $6^3 \times 4$ lattices with staggered fermions~\cite{fk02}.
While the multi-parameter reweighting is successful near the
transition line, it is not clear how to extend it beyond this region,
particularly the $T = 0$ case where one wants to keep the $\beta$ and
quark mass fixed while changing the $\mu$. One still expects to face
the overlap problem in the latter case. It is shown~\cite{aeh02} that
Taylor expanding the observables and the rewriting factor leads to
coefficients expressed in local operators and thus admits study of
larger volumes, albeit still with small $\mu$ at finite temperature.

   In the imaginary chemical potential approach, the fermion determinant is real
and one can avoid the sign problem~\cite{dms90,wei87,akw00,dp02}. In practice,
a reference imaginary chemical potentials is used to carry out the Monte
Carlo calculation and the determinants at other chemical potential values are calculated
through a bosonic Monte Carlo calculation so that one can obtain the finite baryon partition 
function $Z_B(T,V)$ through the Fourier transform of the grand canonical partition function
$Z_{GC} (\mu/T, T, V)$~\cite{akw00}. However. this is problematic for large systems when
the determinant cannot be directly calculated and it still suffers from the
overlap problem. The QCD phase diagram has been studied with  physical observables Taylor expanded 
and analytically continued to the real $\mu$~\cite{dp02}. Again, due to the overlap problem, 
one is limited to small real $\mu$ near the finite temperature phase transition.

\section{Finite Baryon Density -- A Canonical Ensemble Approach}
\label{finite_baryon}

   An algorithm based on the canonical ensemble approach to overcome the 
overlap problem at zero temperature is
proposed~\cite{kfl02}.  To avoid 
the overlap problem, one needs to lock in a definite nonzero baryon sector so that
the exponentially large contamination from the zero-baryon sector is
excluded. To see this, we first note that the fermion
determinant is a superposition of multiple quark loops of all sizes
and shapes. This can be easily seen from the property of the determinant
\begin{equation} 
det M = e^{Tr \log M} = 1 + \sum_{n =1} \frac{(Tr \log M)^n}{n!}.
\end{equation}
Upon a hopping expansion of $\log M$, $Tr \log M$ represents a sum of 
single loops with all sizes and shapes. The determinant is then 
the sum of all multiple loops. The fermion loops can be separated into
two classes. One is those which do not go across the time boundary and
represent virtual quark-antiquark pairs; the other includes those
which wraps around the time boundary which represent external quarks
and antiquarks. The configuration with a baryon number one which contains three
quark loops wrapping around the time boundary will have
an energy $M_B$ higher than that with zero baryon number. Thus, it is
weighted with the probability $e^{- M_B N_t a_t}$ compared with the one 
with no net baryons. We see from the above discussion that the fermion 
determinant contains a superposition of sectors of all baryon numbers, positive, 
negative and zero. At zero temperature where $M_B N_t a_t \gg 1$, the 
zero baryon sector dominates and all the other baryon sectors are exponentially 
suppressed. It is obvious that to avoid the overlap problem, one
needs to select a definite nonzero baryon number sector and stay in it
throughout the Markov chain of updating gauge configurations. To select a
particular baryon sector from the determinant can be achieved by
the following procedure~\cite{fab95}: first, assign an $U(1)$ phase factor 
$e^{-i \phi}$ to the links between the time slices $t$ and $t + 1$ 
so that the link $U/U^{\dagger}$ is multiplied by $e^{-i \phi}/e^{i \phi}$;
then the particle number projection can be carried out through the
Fourier transformation of the fermion determinant like in the BCS theory
\begin{equation}  \label{projection}
P_N = \frac{1}{2 \pi} \int_0^{2 \pi} d\phi e^{-i \phi N} det M[\phi]
\end{equation}
where $N$ is the net quark number, i.e. quark number  minus antiquark number.
Note that all the virtual quark loops which do not reach the time
boundary will have a net phase factor of unity; only those with a 
net N quark loops across the time boundary will have a phase factor
$e^{i \phi N}$ which can contribute to the integral in Eq. (\ref{projection}).
Since QCD in the canonical formulation does not break $Z(3)$
symmetry, it is essential to take care that the ensemble is
canonical with respect to triality. To this end, we shall consider
the triality projection~\cite{fab95,fbm95} to the zero triality sector
\begin{equation}
det_0 M = \frac{1}{3} \sum_{k = 0, \pm 1} det M [\phi + k 2\pi/3].
\end{equation}   
This amounts to limiting the quark number N to a multiple of 3. Thus
the triality zero sector corresponds to baryon sectors with integral
baryon numbers.

\begin{figure}[h]
\includegraphics{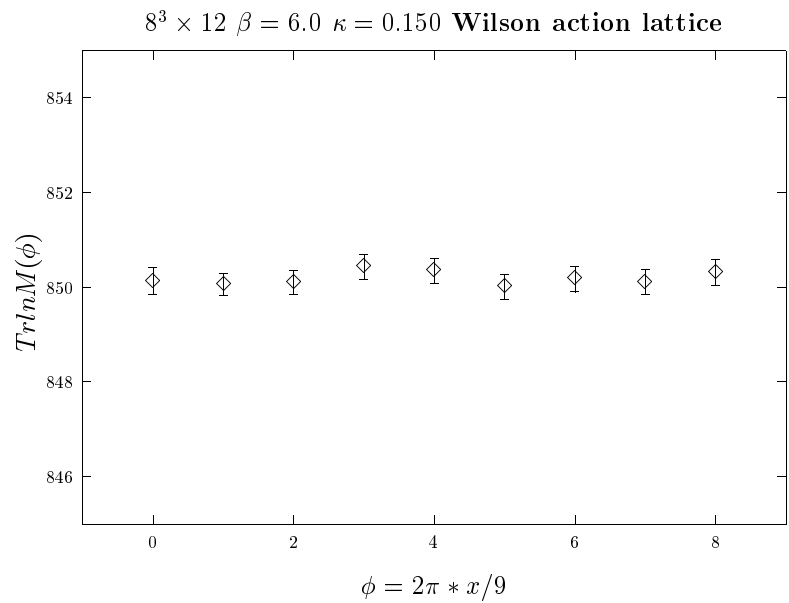}
\vspace{8cm}
\caption{$Tr \log M[\phi]$ for a $8^3 \times 12$ configuration with Wilson
action as a function of $\phi$. }
\end{figure}

\begin{figure}[tb]
\includegraphics{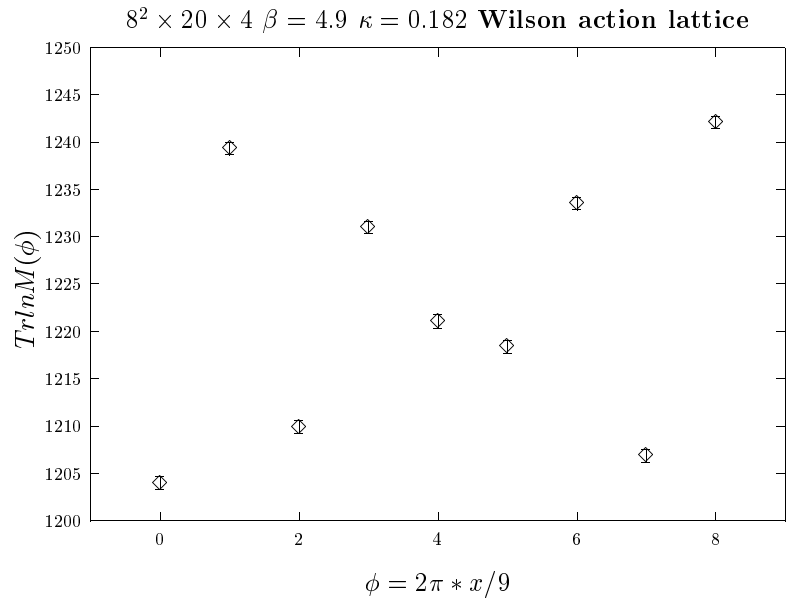}
\vspace{8cm}
\caption{$Tr \log M[\phi]$ for a $8\times 20^2 \times 4$ finite temperature 
configuration with dynamical fermion.}
\end{figure}

 Another essential ingredient to circumvent the overlap problem is
to stay in the chosen nonzero baryon sector so as to avoid mixing
with the zero baryon sector with exponentially large weight.
This can be achieved by preforming the baryon number projection
as described above {\it before} the accept/reject step in the 
Monte Carlo updating of the gauge configuration. If this is not
done, the accepted gauge configuration will be biased toward
the zero baryon sector and it is very difficult to project out
the nonzero baryon sector afterwords. This is analogous to the
situation in the nuclear many-body theory where it is known~\cite{wong75} that the
variation after projection (Zeh-Rouhaninejad-Yoccoz method~\cite{zeh65,ry66}) 
is superior than the variation before projection (Peierls-Yoccoz method~\cite{py57}).
The former gives the correct nuclear mass in the 
case of translation and yields much improved wave functions in mildly deformed 
nuclei than the latter.

   To illustrate the overlap problem, we plot in Fig.1 \,$Tr \log M[\phi]$ for
a configuration of the $8^3 \times 12$ lattice with the Wilson action
at $\beta = 6.0$ and $\kappa = 0.150$ which is obtained with 500 $Z_2$
noises. We see that the it is rather flat in $\phi$ indicating that
the Fourier transform in Eq. (\ref{projection}) will mainly favor the
zero baryon sector. On the other hand, at finite temperature, it is
relatively easier for the quarks to be excited so that the zero baryon
sector does not necessarily dominate other baryon sectors. Another way of seeing this
is that the relative weighting factor $e^{- M_B N_t a_t}$ can be
$O(1)$ at finite temperature. Thus, it should be easier to project out
the nonzero baryon sector from the determinant. We plot in Fig. 2
a similarly obtained $Tr \log M[\phi]$ for configuration of the 
$8 \times 20^2 \times 4$ lattice at finite temperature with $\beta = 4.9$ and 
$\kappa = 0.182$. We see from the figure that there is quite a bit of wiggling in this
case as compared to that in Fig. 1. This implies that it is easier to
project out a nonzero baryon sector through the Fourier transform at finite
temperature.

\section{Noisy Monte Carlo with Fermion Determinant}
\label{nmc}

    In order to implement the canonical ensemble approach, it is clear that
one needs to evaluate the fermion determinant for the purpose of particle
projection. Since the pseudofermion approach does not give the determinant
in the Markov process, it is not applicable. In view of the fact that it is 
impractical to calculate the determinant directly for realistic volumes, 
a Monte Carlo algorithm which accommodates an unbiased estimate of
the probability and an efficient way to estimate the determinant are 
necessary for the finite baryon density calculation. 

    A noisy Monte Carlo algorithm~\cite{lls99} with  Pad\'{e}-Z$_2$ 
estimates~\cite{TDLY,dl94} of the $Tr \log$ of the fermion matrix are 
developed toward this goal and a numerical simulation with Wilson dynamical
fermion is carried out~\cite{jhl03}. We shall summarize the progress made so
far.

    The QCD partition function can be written in the form

\begin{eqnarray}
   Z & = & \int dU \, e^{-S_g(U)}
    \int  \prod_{i=1}^\infty d \eta_{i} \ P^{\eta}(\eta_{i}) \nonumber \\
    &\times &  \prod_{k=2}^\infty d \rho_{k} \ P^{\rho}(\rho_{k})\;
       f(U,\eta,\rho) \, ,
   \label{eq:10}
\end{eqnarray}
where $S_g(U)$ is the gauge action. $f(U, \eta, \rho)$ stands for $f(U, \{ \eta_{i} \}, \{ \rho_k \})$
which is an unbiased stochastic estimator~\cite{bk85}  of the fermion determinant $e^{Tr \log M}$ via 
an infinite number of auxiliary variables $\rho_k$ and $\eta_i$. 
$P^{\eta}(\eta_{i}) = \delta(|\eta_i| -1)$ is the distribution for the $Z_2$ noise $\eta_i$
and $P^{\rho}(\rho_k) = \theta(\rho_k) - \theta(\rho_k-1)$ is the flat distribution for 
$0\le \rho_k \le 1$. With $f(U, \{ \eta_{i} \}, \{ \rho_k \})$ being the stochastic expansion 

\begin{eqnarray}
f(U, \{ \eta_i \}, \{ \rho_k \}) =\, 1 + \left\{ x_1
  + \theta \left(1 - \rho_2\right) \left\{ x_2
  + \theta \left(\frac{1}{3} - \rho_3\right) \left\{ x_3 + \ldots  \right. \right. \right. \nonumber \\
 \ldots \left.  \left. \left. + \theta \left(\frac{1}{n} -\rho_{n} \right) \left\{x_n + \ldots \right\} 
\right\} \right\}  \right\}
\label{e:stochExp}
\end{eqnarray}
where $x_i = \eta^{\dagger}_i \ln M(U)\eta_i$, one can verify~\cite{bk85} that 

\begin{equation}
{\prod_{i=1}^{\infty} d \eta_i P^{\eta}(\eta_i) 
\prod_{k=2}^{\infty} d \rho_k P^{\rho}(\rho_k)} \langle f(U, \{ \eta_{i} \}, \{ \rho_{k} \}) \rangle
= e^{Tr \ln M(U)}, \ 
\end{equation}
and the stochastic series terminates after $e$ terms on the average.

    Since the estimator $f(U,\eta,\rho)$ can be negative due to the stochastic
estimation, the standard treatment is to absorb the sign into the observables, i.e.

\begin{equation} \label{eq:6}
\langle \mathcal{O} \rangle_P  \;=\; 
\frac{\;\langle\, \mathcal{O}\, {\rm sgn}(P)\,\rangle_{|P|}\;}
     {\;\langle\, {\rm sgn}(P) \,\rangle_{|P|}\;}   \ .
\end{equation}

   With the probability for the gauge link variable $U$ and noise $\xi \equiv (\eta, \rho)$ written as 
$P(U,\xi) \propto P_1(U) P_2(U,\xi) P_3(\xi)$ with
\begin{eqnarray} \label{e:P1P2P3Def}
P_1(U)  & \propto & e^{-S_g(U)} \nonumber \\
P_2(U, \xi) & \propto &  | f(U, \xi) | \nonumber \\
P_3(\xi) & \propto & \prod_{i=1}^\infty P^{\eta}(\eta_{i})\,
                 \prod_{k=2}^\infty P^{\rho}(\rho_k),
\end{eqnarray}
the following two steps are needed to prove detailed balance~\cite{lls99,jhl03}.

(a) Let $T_1(U,U')$ be the ergodic Markov matrix satisfying
detailed balance with respect to $P_1$, in other words
$P_1(U) T_1(U,U') dU = P_1(U')T_1(U',U) dU'$. Then the transition matrix
\begin{equation} \label{e:MetropGauge}
    T_{12}(U,U') \;=\; T_1(U,U')\,  
    \min\, \Bigl[ 1,\, {{P_2(U',\xi)}\over{P_2(U,\xi)}}\,\Bigr]
\end{equation}
satisfies detailed balance with respect to the $P_1(U) P_2(U,\xi)$ 
(with $\xi$ fixed).

(b) The transition matrix
\begin{equation} \label{e:MetropNoise}
    T_{23}(\xi,\xi') \;=\; P_3(\xi')\, 
    \min\, \Bigl[ 1,\,{{P_2(U,\xi')}\over{P_2(U,\xi)}}\,\Bigr]
\end{equation}
satisfies detailed balance with respect to $P_2(U,\xi) P_3(\xi)$
(with $U$ fixed).

From (a), (b) it follows that $T_{12}$ and $T_{23}$ keep the original 
distribution $P(U,\xi)$ invariant and interleaving them will lead to
an ergodic Markov process with the desired fixed point.

\subsection{Pad\'{e} - Z$_2$ Estimator of $Tr \ln M$ with Unbiased Subtraction}

    In Eq. (\ref{e:stochExp}), one needs to calculate $x_i = \eta^{\dagger}_i \ln M(U)\eta_i$
in the stochastic series expansion of the fermion determinant. An efficient method is
developed to calculate it~\cite{TDLY}. First of all, the logarithm is approximated 
using a Pad\'e approximation, which after the partial fraction expansion, 
has the form
\begin{equation} \label{e:Pade}
\ln M(U, \kappa) \approx R_{M}(U) \equiv  b_0 \ I + \sum_{i=1}^{N_{P}} 
b_{i}\left(M(U, \kappa) + c_{i} \ I \right)^{-1}
\end{equation}
where $N_{P}$ is the order of the Pad\'e approximation, and the constants 
$b_{i}$ and $c_{i}$ are the Pad\'e coefficients. In our implementation 
we have used an 11-th order approximation whose coefficients are 
tabulated in \cite{TDLY}. The traces of $\ln M$ are then estimated
by evaluating bilinears of the form $\eta^{\dagger} R_{M}(U) \eta$. If the
components of $\eta$ are chosen from the complex $Z_2$
group, then the contributions to the variance of these bilinears come only from
off diagonal elements of $R_M(U)$~\cite{bmt94,dl94}. 
In this sense, $Z_2$ noise is optimal and has been applied to the calculation of
nucleon matrix elements involving quark loops~\cite{dll95}.  An effective method 
reducing the variance is to subtract off a linear combination
of traceless operators from $R_{M}(U)$ and to consider
\begin{equation}
E[Tr \ R_{M}(U), \eta ] = \eta^{\dagger} \left( R_{M}(U) - \alpha_{i} \mathcal{O}_{i} \right) \eta  \ .
\end{equation}
 Here the $\mathcal{O}_{i}$ are
operators with $Tr \ \mathcal{O}_{i} = 0$.  Clearly since the $\mathcal{O}_{i}$ are
traceless they do not bias the estimators. The $\alpha_{i}$
are constants that can be tuned  to minimize the fluctuations in $E[Tr \ R_{M}(U), \eta]$.

With other types of noise 
such as Gaussian noise, the variance receives contributions from diagonal terms which one
cannot subtract off. In this case, the unbiased subtraction scheme
described here is ineffective. In practice, the $\mathcal{O}_{i}$ are constructed by taking 
traceless terms from the hopping parameter expansion for $M^{-1}(U)$. It is shown
for Wilson fermions on a $8^3 \times 12$ lattice at $\beta = 5.6$, these subtractions can reduce the
noise coming from the terms $\left( M(U) + c_{i} \right)^{-1}$ in equation
(\ref{e:Pade}) by a factor as large as 37 for $\kappa = 0.150$ with 50 Z$_2$ noises~\cite{TDLY}.

\subsection{Implementation of the Noisy Monte Carlo Algorithm}

    The noisy Monte Carlo algorithm has been implemented for the Wilson dynamical
fermion with pure gauge update (Kentucky Noisy Monte Carlo Algorithm) for an
$8^4$ lattice with $\beta = 5.5$ and $\kappa = 0.155$~\cite{jhl03}. Several tricks
are employed to reduce the fluctuations of the $Tr \ln M$ estimate and increase the
acceptance. These include shifting the $Tr \ln M$ with a constant, $\Delta \beta$
shift~\cite{sw95}, and splitting the $Tr \ln M$ with $N$ `fractional flavors'.
After all these efforts, the results are shown to agree with those from the  HMC 
simulation. However,
the autocorrelation is very long and the acceptance rate is low. This has to do
with the fact that $Tr \ln M$ is an extensive quantity which is proportional to
volume and the stochastic series expansion of $e^x$ converges for $x \le 6$ for
a sample with the size of $\sim 10^3 - 10^4$. This is a stringent requirement
which requires the fractional flavor number $N \ge 15$ for this lattice. This can
be seen from the distribution of $x = \sum_f (Tr R_M^f (U) - \lambda^f Plag - x_0^f)/N$
in Fig. 3 which shows that taking $N$ to be 15, 20, and 25, the largest $x$ value
is less than 6.

\begin{figure}
\includegraphics[width=3.3in]{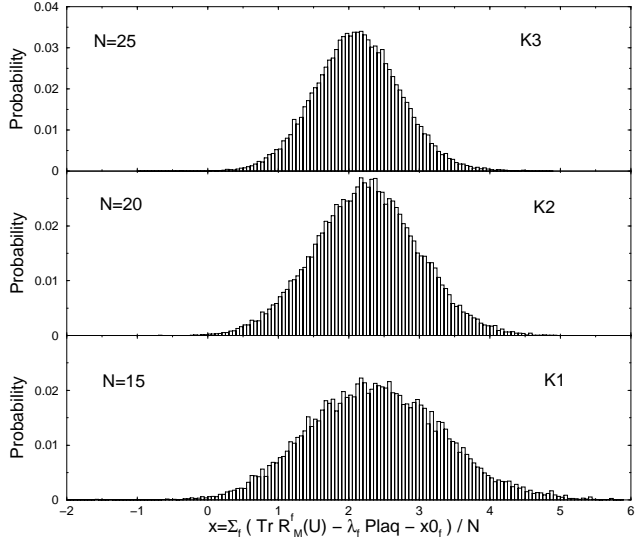}
\caption{\label{f:TrLnDist}Distributions of $x$ for the three noisy simulations}
\end{figure}

  As the volume increases, this fractional flavor needs to be larger to keep
$x$ smaller than 6 for a sample of the size $\sim 10^3 - 10^4$. At the 
present volume ($8^4$), the autocorrelation is already much longer than that
of HMC, it is going to be even less efficient for larger volumes. This is 
generic for the noisy Monte Carlo algorithm which scale with volume as $V^2$, while
HMC scales as $V^{5/4}$.

\section{Hybrid Noisy Monte Carlo Algorithm -- a New Proposal}

   It is clear that the inefficiency of the noisy Monte Carlo algorithm for the
fermion determinant is due to the large fluctuation of the $Tr \ln M$ estimator
from one gauge configuration to the next. We shall propose a combined Hybrid Monte 
Carlo (HMC) and Noisy Monte Carlo (NMC) to remove such fluctuations
in the context of the finite density. 

   With the baryon number projection discussed in Sec. \ref{finite_baryon}, we can
write the partition function for the finite baryon sector with $B$ baryons as

\begin{equation}
Z_B = \int dp dU d \phi^{\dagger} d\phi e^{-p^2/2 - S_g (U) + \phi^{\dagger}
          (M^{\dagger}M)^{-1}\phi} \frac{ \frac{1}{2 \pi} \int_0^{2 \pi} d\theta e^{-i 3 B\theta} 
        det M^{\dagger}M[\theta]}{det M^{\dagger}M[\theta = 0]}.
\end{equation}
In this case, one can update the momentum $p$, the gauge link variable $U$ and the
pseudofermion field $\phi$ via HMC and then interleave with NMC for updating 
the determinant ratio 
\begin{equation}  \label{R}
R = \frac{ \frac{1}{2 \pi} \int_0^{2 \pi} d\theta e^{-i 3 B\theta} 
        det M^{\dagger}M[\theta]}{det M^{\dagger}M[\theta = 0]}.
\end{equation}

As described in Sec. \ref{nmc}, NMC involves two Metropolis accept/reject steps to 
update the ratio with the Pad\'{e} - Z$_2$ estimator of the $Tr \ln$ difference of the
determinants, i.e. $Tr (\ln M^{\dagger}M[\theta] - \ln  M^{\dagger}M[\theta = 0])$.
It is pointed out~\cite{fab96} that for zero temperature, one can approximate the continuous
integral over $\theta$ with a discrete sum incorporating triality zero 
projection~\cite{fab95,fbm95} so that the partition function is a mixture
of different $Z_B$ for different baryon number B. In other words, the
approximation  
\begin{equation} \label{projection_appro}
\frac{1}{2 \pi} \int_0^{2 \pi} d\theta e^{-i 3 B\theta} 
        det M^{\dagger}M[\theta]  \longrightarrow 
\frac{1}{3 B_N} \sum_{k =0}^{3 B_N -1} e^{ -i \frac{2\pi k B}{3 B_N}} 
det  M^{\dagger}M[\frac{2\pi k B}{3 B_N}]
\end{equation}
leads to the mixing of the baryon sector $B$  with those of $B \pm B_N, B \pm 2 B_N...$.
If $B$ is small and $N_B > B$, then the partition will be dominated by
$Z_B$ with small mixture from $Z_{B \pm B_N}, Z_{B \pm 2 B_N},...$.
For example, if we take $B = 1$ and $B_N = 5$,  the discrete approximation
gives an admixture of partition function with baryon number $B =1, 5, 11, -4, -9, ...$.
At zero temperature, the partition function $Z_B$ behaves like 
$e^{- B m_N N_t a_t}$, one expects that the mixing due to baryons other than $B = 1$
will be exponentially suppressed when $m_n N_t a_t > 1$.

    Two points need to be stressed. First of all, it is crucial to
project out the particle number in Eq. (\ref{projection_appro}) before the Metropolis
accept/reject step in order to overcome the overlap problem. Secondly, given that
the ratio $R$ in Eq. (\ref{R}) is replaced with a discrete sum

\begin{equation}
\overline{R} = \frac{1}{3 B_N} \sum_{k =0}^{3 B_N -1} e^{ -i \frac{2\pi k B}{3 B_N}}
e^{Tr (\ln M^{\dagger}M[\frac{2\pi k B}{3 B_N}] - \ln  M^{\dagger}M[0])},
\end{equation}
which involves the difference between the $Tr \ln M^{\dagger}M[\frac{2\pi k B}{3 B_N}]$ 
and $Tr  \ln  M^{\dagger}M[0]$, it takes out the fluctuation due to the gauge configuration
which plagued the Kentucky Noisy Monte Carlo simulation in Sec. \ref{nmc}.
Furthermore, the $Tr \ln$ difference is expected to be $O(1)$ as seen from
Fig. 1. If indeed is the case, it should lead to a better convergence of the stochastic 
series expansion in Eq. (\ref{e:stochExp}) and the algorithm scales with volume the same
as HMC.

\subsection{Another Application}

  Here we consider another possible application of the Hybrid Noisy Monte Carlo algorithm.
HMC usually deals with two degenerate flavors. However, nature comes with 3 distinct
light flavors -- $u,d$ and $s$. To consider $u$ and $d$ as separate flavors, one
can perform HMC with two degenerate flavors at the $d$ quark mass and then employ NMC 
to update the determinant ratio

\begin{equation}  \label{R_ud}
R_{ud} = \frac{Det M^{\dagger}_d Det M_u}{Det M^{\dagger}_d Det M_d}
 = e^{Tr (\ln M_u - \ln M_d)}.
\end{equation}
Since both the $u$ and $d$ masses are much smaller than $\Lambda_{QCD}$, 
$Tr (\ln M_u - \ln M_d)$ should be small. If the $Tr \ln$ difference is
small enough (e.g. $O(1)$) so that the acceptance rate is high, it could be
a feasible algorithm for treating $u$ and $d$ as distinct flavors so that  
realistic comparison with experiments can be done someday. It is shown recently that the
Rational Hybrid Monte Carlo Algorithm (RHMC)~\cite{hks99,ck03} works efficiently for 
two flavor staggered fermions. It can be applied to single flavors for 
Wilson, domain wall, or overlap fermions at the cost of one pesudofermion
for each flavor. We should point out that, in comparison, the  Hybrid Noisy approach 
discussed here saves one pseudofermion, but at the cost of having to update the
determinant ratio $R_{ud}$ in Eq. (\ref{R_ud}).

  While we think that the Hybrid Noisy Monte Carlo algorithm proposed here
might overcome the overlap and the low acceptance problems
and the determinant $det M[\theta]$ is real in this approach, 
the fact that the Fourier transform in 
Eq. (\ref{projection_appro}) involves the baryon number $B$ 
may still lead to a sign problem
in the thermodynamic limit when $B$ and $V$ are large. 
However, as an initial attempt, we are more interested in finding out if the 
algorithm works for a small $B$ such as 1 or 2  in a relatively small box.

\section{Acknowledgment}
   This work is partially supported by  DOE grants DE-FG05-84ER0154 and
DE-FG02-02ER45967. The author would like to thank the organizers of
the ``Third International Workshop on QCD and Numerical Analysis' for the
invitation to attend this stimulating workshop. He also thanks B. Jo\'{o},
S.J.Dong, I. Horv\'{a}th, and M. Faber for stimulating discussions of
the subject.



\printindex
\end{document}